\documentclass[aps,prl,twocolumn,showpacs]{revtex4}

\usepackage{amsmath}
\usepackage{epsfig}
\newcommand{\bea}{\begin{eqnarray}}
\newcommand{\eea}{\end{eqnarray}}
\newcommand{\be}{\begin{equation}}
\newcommand{\ee}{\end{equation}}

\begin{document}

\title{Fermi-edge singularity in a spin-incoherent Luttinger liquid}

\author{Gregory A. Fiete}

\affiliation{Kavli Institute for Theoretical Physics, University of California, Santa Barbara, California 93106, USA}

\begin{abstract}

We theoretically investigate the Fermi edge singularity in a spin incoherent Luttinger liquid.  Both cases of finite and infinite core hole mass are explored, as well as the effect of a static external magnetic field of arbitrary strength. For a finite mass core hole the absorption edge behaves as $(\omega-\omega_{\rm th})^\alpha/ \sqrt{|\ln(\omega-\omega_{\rm th})|}$ for frequencies $\omega$ just above the threshold frequency $\omega_{\rm th}$. The exponent $\alpha$ depends on the interaction parameter $g$ of the interacting one dimensional system, the electron-hole coupling,  and is independent of the magnetic field strength, the momentum, and the mass of the excited core hole (in contrast to the spin coherent case).   In the infinite mass limit, the spin incoherent problem can be mapped onto an equivalent problem in a spinless Luttinger liquid for which the logarithmic factor is absent, and backscattering from the core hole leads to a universal contribution to the exponent $\alpha$.  

\end{abstract}

\date{\today}
\maketitle



One dimensional (1-d) interacting systems have received a great deal of theoretical and experimental attention.  In part, this is because 1-d systems exhibit exceptionally rich physics, the canonical example of which is the Luttinger liquid (LL).  The LL displays power law correlations with exponents that depend on the interactions and the symmetries of the problem [\onlinecite{Voit:rpp95,Haldane81}].  In the case of a gas of fermions, ``exotic'' effects such as spin-charge separation may also be realized in a LL. The overwhelming experimental evidence that such states of matter do exist in realizable physical situations has further stimulated and challenged theory [\onlinecite{ishii:nat03,Bockrath:nat99,Yao:nat99,Auslaender:sci02,Auslaender:sci05}]. 

Recently, the ``spin incoherent Luttinger liquid'' (SILL) has received much attention because of its distinct properties that partially resemble those of a LL but partially do not [\onlinecite{cheianov03,Cheianov04,Fiete:prl04,Fiete:prb05,Matveev:prl04,Matveev:prb04,Fiete_2:prb05,Fiete:prb06,Kindermann_noise:prb06,Kindermann:prl06,Kakashvili:cm06,Fiete:rmp07}].  The crucial difference between a LL and a SILL is the {\em combination} of (i) very strong particle-particle interaction and (ii) finite but small temperature.  In a SILL the charge degrees of freedom are only very weakly influenced by the temperature, while the spin degrees of freedom are highly thermally excited. For the purposes of this paper, we will discuss a 1-d electron gas, but the physics of the SILL is the same for a 1-d hole gas.  The role of  strong interactions in the SILL is to drive a large (possibly exponentially large) separation between spin and charge energy scales. For an infinite system, the generic form of the spin-charge separated Hamiltonian (for energies small compared to the charge scale, but arbitrary compared to the spin scale) is $H_{\rm elec}=H_c + H_s+H_Z$, where
\be
\label{eq:H_c}
H_c=\hbar v \int \frac{dx}{2\pi}\left[\frac{1}{g} (\partial_x\theta(x))^2+g(\partial_x\phi(x))^2\right],
\ee
and
\be
\label{eq:H_s}
H_s= J \sum_l {\vec S_l}\cdot {\vec S_{l+1}},\;\; H_Z=-g_e\mu_B B \sum_l S^z_l.
\end{equation}
The Hamiltonian \eqref{eq:H_c} describes the low-energy density fluctuations of the electron gas. The energy scale for the charge sector is set by $E_c \sim \hbar v/a$ where $\hbar$ is Planck's constant, $v$ is the collective mode velocity and $a$ is the average spacing between electrons.  The parameter $g$ describes the strength of the microscopic interactions and the bosonic fields appearing in Eq.~\eqref{eq:H_c} satisfy the commutation relations $[\partial_x\theta(x),\phi(x')]=i\pi \delta(x-x')$.  For strong interactions (which typically occur at very low electron density) the spin degrees of freedom behave like a 1-d antiferromagnetic spin chain. In Eq.~\eqref{eq:H_s}, ${\vec S_l}$ is the spin of the $l^{th}$ electron,    $J>0$ is the nearest neighbor exchange energy, and $g_e \mu_B$ is the effective coupling of the external magnetic field $B$.  The two central points to be taken from \eqref{eq:H_c} and \eqref{eq:H_s} is that the ratio $J/E_c$ can be (and will be assumed here to be) exponentially small and the charge sector is described by harmonic fluctuations over the relevant energy scales.  The actual form of the spin Hamiltonian \eqref{eq:H_s} is immaterial for our discussion. It is only introduced to provide a concrete reference for the energy scale $J$.

Besides the vanishing ratio $J/E_c$, the other necessary ingredient for the SILL is finite temperature. Specifically, we are interested in temperatures such that $J \ll k_B T \ll E_c$, where $k_B$ is Boltzmann's constant.  All our calculations are done in the limit $J\to 0$, then $T\to 0$.  As a result, in the evaluation of any correlation function the Boltzmann factors for all spin states will be identical (aside from trivial Zeeman factors) as the limit described above is equivalent to $e^{-\beta H_s}\equiv1$ (with $\beta=(k_B T)^{-1})$ which is why the precise form of the spin Hamiltonian \eqref{eq:H_s} is unimportant in the SILL. One may say that the SILL exhibits ``super universal'' spin physics.

\begin{figure}[t]
\includegraphics[width=.8\linewidth,clip=]{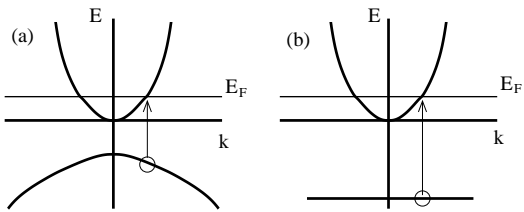}
\caption{\label{fig:absorption} Schematic of threshold photo-excitation for (a) finite hole mass and (b) infinite hole mass.  In (a) an electron is excited from the valence band to the conduction band leaving behind a hole in valence band.  In (b) an electron is excited from a deep (infinite mass) core level.  In this paper we are assuming the ``conduction band" is occupied by a SILL so the threshold transitions shown are actually to $2k_F$ (rather than $k_F$)[\onlinecite{Fiete:prl04}].}
\end{figure}

While many properties (such as those related to particle-conserving operators[\onlinecite{Fiete_2:prb05}]) of the SILL can be mapped onto a spinless LL, the single-particle Greens function exhibits qualitatively unique behavior and does not appear to correspond to any unitary conformal field theory[\onlinecite{cheianov03,Cheianov04,Fiete:prl04}].
Implications for momentum resolved tunneling[\onlinecite{Fiete:prb05}], transport[\onlinecite{Matveev:prl04,Matveev:prb04,Fiete_2:prb05,Kindermann_noise:prb06,Kindermann:prl06}], and Coulomb drag[\onlinecite{Fiete:prb06}] have been worked out.  Preliminary experimental indications point to the realization of the SILL in quantum wires with low electron density[\onlinecite{Steinberg:prb06}].  In this Letter we are concerned with the Fermi-edge singularity[\onlinecite{Ogawa:prl92,Balents:prb00,Kane:prb94,Calleja:prb95}]  and its magnetic field dependence in a SILL.  See Fig.~\ref{fig:absorption} for a schematic. The photon absorption is given by
\be 
\label{eq:threshold}
I(\omega) \propto \sum_\sigma {\rm Re}\int_0^\infty dt e^{i\omega t} \langle \psi_\sigma(t)h_\sigma(t)h_\sigma^\dagger(0)\psi_\sigma^\dagger(0)\rangle,
\ee
where the operator $\psi_\sigma^\dagger$ ($\psi_\sigma$) creates (annihilates) an electron and $h_\sigma^\dagger$ ($h_\sigma$) creates (annihilates) a hole with spin $\sigma$.  There will be some threshold frequency $\omega_{\rm th}$ below which there is no absorption, $I(\omega) \equiv 0$ for $\omega < \omega_{\rm th}$, while for $\omega > \omega_{\rm th}$,  $I(\omega)\sim (\omega -\omega_{\rm th})^\alpha f(\omega)$. The main result of this work is that for the SILL with a finite hole mass, $\alpha=(1-\delta_s)^2/(2g)-1$ and $f(\omega)=1/\sqrt{|\ln(\omega-\omega_{\rm th})|}$ where $\delta_s$ and $\omega_{\rm th}$ are given below. In the case of an infinite hole mass, $\alpha=(1-\delta_s)^2/g+1/8-1$ and $f(\omega)=1$.  Here $\delta_s$ is a forward scattering contribution to the exponent $\alpha$ and in the case of an infinitely massive hole, there is a universal backscattering contribution of $1/8$ to $\alpha$[\onlinecite{Furusaki:prb97,Komnik:prb97,Gogolin:prl93,Prokofev:prb94}].  At frequencies just above $\omega_{\rm th}$ these results are independent of the magnetic field (provided it is not infinite).

The important distinction between finite and infinite hole mass is that the latter breaks the translational symmetry of the system while the former does not.  The most significant consequence of this is that backscattering from an infinitely massive impurity is relevant and cuts the electron system into two semi-infinite parts [\onlinecite{Kane:prl92,Furusaki:prb93}], while backscattering from a finite mass impurity is irrelevant[\onlinecite{Neto:prb96}].  These two limits have important implications for the boundary conditions the $\theta$ and $\phi$ operators in Eq.~\eqref{eq:H_c} satisfy in each case, which in turn effects  the Fermi-edge singularity since $\theta$ and $\phi$ appear in the $\psi_\sigma$ (see below) in Eq.~\eqref{eq:threshold}. We first turn to the more interesting case of a finite hole mass.

In the case of a finite hole mass, we follow the procedure of Refs.~[\onlinecite{Neto:prb96,Tsukamoto:prb98,Tsukamoto:epj98}] and transform to a frame comoving with the excited hole. In this frame the Hamiltonian takes the form $H=H_{\rm elec}+H_{\rm elec-hole}+H_{\rm hole}$, where
\be
H_{\rm elec-hole}=\frac{U_s^f}{\pi} h^\dagger h \partial_x\theta(0) \pm  \frac{U_a^f}{\pi}h^\dagger h \partial_x\phi(0)\,,
\ee
and  $H_{\rm hole}=\sum_\sigma E_{h,\sigma} h_\sigma^\dagger h_\sigma$ and $h^\dagger h=\sum_\sigma h_\sigma^\dagger h_\sigma$.  Here $U_s^f$ is the symmetric part of the forward scattering from the hole and $U_a^f$ is the antisymmetric part of the forward scattering [\onlinecite{Tsukamoto:prb98}]. 
(In our convention $\partial_x\theta$ represents the density fluctuations and $\partial_x\phi$ the particle current.)  The antisymmetric part appears since in the frame of the hole, it sees a net current of particles scattering from it. The ``+'' sign is for a right-moving hole and the ``-'' sign is for a left-moving hole. The parameter $U_a^f$ depends on the momentum and mass of the hole, and when it is at rest,  $U_a^f\equiv 0$[\onlinecite{Tsukamoto:prb98}].  For the remainder of this paper, we will only consider a right moving hole, as shown in Fig.~\ref{fig:absorption}, as the left moving contribution is identical.  Since the backscattering from a finite mass impurity is not relevant it has no affect on the Fermi-edge physics and therefore has not been included in $H_{\rm elec-hole}$.

The Hamiltonian $H$ can be diagonalized with the unitary transformation $U={\rm exp}\{-i[\delta_a\theta(0)+\delta_s\phi(0)]h^\dagger h\}$ where $\delta_a\equiv \mp U_a^f/(vg\pi)$ and $\delta_s \equiv -gU_s^f/(v\pi)$.
Applying this transformation we find $\bar H  \equiv U^\dagger H U =  H_{\rm elec}+\bar H_{\rm hole}$, where the only change to $H_{\rm hole}$ is a shift in the hole energy $E_{h,\sigma} \to \tilde E_{h,\sigma}$, which is unimportant to us here.  Our main concern is with computing the threshold exponent $\alpha$ and the form of the function $f(\omega)$. (See discussion below Eq.~\eqref{eq:threshold}.)  We are also interested in the effects of a static external magnetic field.  For a spin coherent LL the Fermi-edge singularity in a finite magnetic field has been addressed in Ref.[\onlinecite{Otani:prb96}].  In a SILL the spin and charge modes are decoupled even for fields strong enough to completely polarized the electrons.  As a result, the magnetic field only appears as a Zeeman energy in the SILL.  In the valence bands, the magnetic field dependence of $\tilde E_{h,\sigma}$ is more subtle since the bands themselves may change in a strong magnetic field. (Different g-factors for holes and electrons, and spin-orbit effects can also complicate the field dependence.)  We do not address this issue in any detail here, but note that the $\tilde E_{h,\sigma}$ will depend on the magnetic field.   In any case, there will always be one particular spin orientation for which the absorption initially occurs and our results below are relevant to this edge.

We begin with the evaluation of the correlation function $C_\sigma(\tau) =\langle \psi_\sigma(\tau)h_\sigma(\tau)h_\sigma^\dagger(0)\psi_\sigma^\dagger(0)\rangle$, where $\tau$ is the imaginary time.  At finite temperature, $C_\sigma(\tau)=\frac{1}{Z}{\rm Tr}\left[e^{-\beta H}  \psi_\sigma(\tau)h_\sigma(\tau)h_\sigma^\dagger(0)\psi_\sigma^\dagger(0)\right]
=\frac{1}{Z}{\rm Tr}\left[e^{-\beta \bar H}  \bar \psi_\sigma(\tau)\bar h_\sigma(\tau)\bar h_\sigma^\dagger(0)\bar \psi_\sigma^\dagger(0)\right],$
where in the second line we have introduced the unitary transformation $U^\dagger U$ and used the cyclic property of the trace. We have already determined $\bar H$, so it remains to determine $\bar \psi_\sigma$ and $\bar h_\sigma$.   Direct evaluation gives $\bar \psi_\sigma = \psi_\sigma$ (up to unimportant multiplicative factors) and $\bar h_\sigma=h_\sigma e^{-i[\delta_a\theta+\delta_s\phi]} $.  Therefore, the correlation function $C_\sigma(\tau)$ separates as $C_\sigma(\tau) = C_{\psi,\sigma}(\tau)C_{h,\sigma}(\tau)$.  One readily finds $C_{h,\sigma}(\tau)=e^{-\tilde E_{h,\sigma} \tau}$ at zero temperature.  This factor will enter the threshold frequency $\omega_{\rm th}$ in Eq.~\eqref{eq:threshold} and will in general also depend on the external magnetic field.  Our main interest here is in the evaluation of the part of the correlation function that will give us the frequency dependence just above threshold,
\bea
\label{eq:C_psi}
C_{\psi,\sigma}(\tau)=\frac{1}{Z_{\rm elec}}{\rm Tr}\biggl[e^{-\beta H_{\rm elec}}   \psi_\sigma(\tau) e^{-i[\delta_a\theta(\tau)+\delta_s\phi(\tau)]}  \nonumber \\
\times e^{i[\delta_a\theta(0)+\delta_s\phi(0)]} \psi_\sigma^\dagger(0)\biggr],
\eea
where $Z_{\rm elec}={\rm Tr}[e^{-\beta H_{\rm elec}}]$.  Formally, Eq.~\eqref{eq:C_psi} bears a striking resemblance to the single particle Green's function evaluated in the spin incoherent regime in Ref.~[\onlinecite{Fiete:prl04}].  It is worth pausing a moment to understand the physics of Eq.~\eqref{eq:C_psi}.  Recalling that the operator $e^{-i \phi(x,\tau)}$ creates a particle at space-time point $(x,\tau)$ in the many-body system and the electron number is related to the $\theta$ field via $N(x,\tau)=\bar nx+\frac{1}{\pi}(\theta(x,\tau)-\theta(0,0))$, with $\bar n$ the average particle density, we see immediately that $C_{\psi,_\sigma}(\tau)$ involves adding an electron {\em plus} an additional ``background excitation" (from $e^{-i\delta_s \phi(0)}$) and then removing  the same particle and its additional background at a time $\tau$ later. The factors $e^{\pm i \delta_a \theta}$ contribute additional density fluctuations coming from the motion of the finite mass valence hole.  This factor is absent in the infinite mass limit. 

The evaluation of the trace in  Eq.~\eqref{eq:C_psi} can be carried out using the  method of
Ref.[\onlinecite{Fiete:prl04}].  Up to unimportant prefactors, $C_{\psi,\sigma}(\tau)$ is given by
\bea
\label{eq:C_psi_s}
C_{\psi,\sigma}(\tau) \sim \sum_m \langle p^\sigma_m (-1)^m\delta(N(\tau)-m)\nonumber \\
\times e^{-i\delta_a[\theta(\tau)-\theta(0)]} e^{i(1-\delta_s)[\phi(\tau)-\phi(0)]}\rangle e^{-E_Z (1-\sigma) \tau/2},
\eea
where $p^\sigma_m$ is the probability of having $m$ electrons parallel with spin $\sigma$, and $(-1)^m$ is a statistical factor coming from permuting fermions. The probability $p^\sigma_m=w_\sigma^{|m|+1}$ where $w_\uparrow=(1+{\rm exp}\{-E_Z/k_BT\})^{-1}$, $w_\downarrow=1-w_\downarrow$, and $E_Z=2g_e\mu_B  B$ is the Zeeman energy of an electron in a magnetic field referenced to the minimum energy configuration with the spin parallel to the field. The remaining trace in \eqref{eq:C_psi_s} over the charge degrees of freedom \eqref{eq:H_c} is represented by $\langle ...\rangle$.

For arbitrary magnetic field strength, $C_{\psi,\sigma}(\tau)$ can be evaluated in closed form by changing $\sum_m \to \int dm$, expressing the delta function as an integral over $\lambda$, and substituting the expression for $p^\sigma_m$
\bea
 C_{\psi,\sigma}(\tau) \sim  w_\sigma \int_{-\infty}^{\infty} \!\!\!\!\! dm \int_{-\infty}^{\infty} \!\!\frac{d\lambda}{2\pi} \langle w_\sigma^{|m|} (-1)^me^{i\lambda(N(\tau)-m)}\nonumber \\
 \times e^{-i\delta_a[\theta(\tau)-\theta(0)]} e^{i(1-\delta_s)[\phi(\tau)-\phi(0)] }\rangle.
\eea
For arbitrary $m$ values, one takes $(-1)^m=(e^{i\pi m}+e^{-i\pi m})/2$ and performs the integration over $m$. 
Expressing $C_{\psi,\sigma}=C^R_{\psi,\sigma}+C^L_{\psi,\sigma}$, 
\bea
\label{eq:C_psi_pm}
C^R_{\psi,\sigma}(\tau) \sim  w_\sigma \int_{-\infty}^{\infty} \frac{d\lambda}{2\pi} \frac{2\ln(1/w_\sigma)e^{-E_Z (1-\sigma) \tau/2}}{(\ln(1/w_\sigma))^2+(\lambda - \pi)^2} \nonumber \\
\times \langle e^{i\lambda N(\tau)}  e^{-i\delta_a[\theta(\tau)-\theta(0)]} e^{i(1-\delta_s)[\phi(\tau)-\phi(0)] }\rangle,
\eea
where it is evident that $C^R_{\psi,\sigma}=C^L_{\psi,\sigma}$ when the correct sign of $\delta_a$ is chosen.
We first note that \eqref{eq:C_psi_pm}  recovers the expected result (a spinless LL) for fully polarized electrons: $w_\uparrow \to 1$ and $w_\downarrow \to 0$ if we recall that $\delta(\tilde \lambda)=\lim_{\epsilon\to 0}\frac{1}{\pi}\frac{\epsilon}{\epsilon^2+\tilde \lambda^2}$. In the present case  $\ln(1/w_\uparrow)$ plays the role of $\epsilon$, so that $C^R_{\psi,\downarrow}=0$, $C^R_{\psi,\uparrow}= \langle e^{i(1-\delta_a)[\theta(\tau)-\theta(0)]} e^{i(1-\delta_s)[\phi(\tau)-\phi(0)] }\rangle$ (neglecting the Zeeman factor) leading to
$C^R_{\psi,\uparrow}(\tau) \sim \left(\frac{\alpha_c}{v\tau}\right)^{\frac{1}{2}\left[g\left(1-\delta_a\right)^2+\frac{1}{g}\left(1-\delta_s\right)^2\right]},$
in agreement with the result obtained in Ref.[\onlinecite{Tsukamoto:prb98}]. 
Here $\alpha_c$ is a short distance cut off of order the lattice spacing $a$. 
To obtain the spinless result, it should be emphasized that the limit $w_\uparrow \to 1$,  $w_\downarrow \to 0$ has first been taken, and then $\tau \to \infty$. This order of limits ensures the long time (low frequency) behavior is governed by the spinless LL where the spins are frozen into a ferromagnetic state by the external field.

For any finite field strength the low frequency behavior is ultimately determined by spin incoherent effects as we will now show. Computing the expectation value in \eqref{eq:C_psi_pm} gives
\bea
\label{eq:C_psi_pm_0}
C^R_{\psi,\sigma}(\tau) \sim \left(\frac{\alpha_c}{v\tau}\right)^{\frac{1}{2g}(1-\delta_s)^2} w_\sigma \int_{-\infty}^{\infty} \frac{d\lambda}{2\pi} \nonumber \\
\times \frac{2\ln(1/w_\sigma)e^{-\frac{(\lambda/\pi-\delta_a)^2}{2} g\ln[v\tau/\alpha_c]}}{(\ln(1/w_\sigma))^2+(\lambda - \pi)^2}e^{-E_Z (1-\sigma) \tau/2}.
\eea
In the extreme long time limit $\tau \to \infty$ the exponential acts as a delta function [$\delta(x)=\lim_{\epsilon\to0} e^{-x^2/\epsilon}\sqrt{\frac{1}{\epsilon\pi}}$, where here $1/\epsilon= \frac{g }{2\pi^2} \ln(v\tau/\alpha_c)$] giving,
upon substitution into \eqref{eq:threshold},  the central result of this paper 
\be
\label{eq:I_SI}
I(\omega) \propto \sum_\sigma (\omega-\omega_{\rm th})^{\frac{1}{2g}(1-\delta_s)^2-1}/\sqrt{|\ln(\omega-\omega_{\rm th})|}
\Theta(\omega-\omega_{\rm th}),
\ee
where $\Theta(\omega-\omega_{\rm th})$ is the step function and $\omega_{\rm th}=\tilde E_{h,\sigma}+E_Z (1-\sigma)/2$.  A few of the most important features of \eqref{eq:I_SI} are worth emphasizing.  Firstly, in contrast to the spin coherent (spin polarized) LL[\onlinecite{Tsukamoto:prb98,Tsukamoto:epj98}] the threshold exponent {\em does not depend on the mass of the core hole}.  The effects of the mass of the core hole only appear in the prefactor.  Secondly, there are ``universal" (independent of interactions and field strength) logarithmic corrections to the power-law threshold behavior.  Thirdly, while the threshold energy $\omega_{\rm th}$ depends on the magnetic field, the exponent $\alpha=\frac{1}{2g}(1-\delta_s)^2-1$ does not.  It is also worth noting that for arbitrary field strength there is a crossover in $I(\omega)$ as a function of $\tilde \omega\equiv \omega-\omega_{\rm th}>0$ from the spin incoherent regime at the smallest $\tilde \omega$ to the ``spinless regime" at larger $\tilde \omega$ given by $\omega^*\sim \omega_c e^{-2E_Z/(gk_BT)}$ with 
$\omega_c=v/\alpha_c$.  Since $1>w_\uparrow >w_\downarrow>0$, the scale $\omega^*$ below which spin incoherent effects appear is set by $w_\uparrow$.   For arbitrary magnetic field strength, in the limit $T \ll \tilde \omega \ll \omega^*\ll \omega_c$, the result \eqref{eq:I_SI} is obtained, while in the limit $T \ll \omega^* \ll \tilde \omega \ll \omega_c$ the spin polarized result is obtained: $I(\omega)\propto \tilde \omega^{\frac{1}{2}\left[g(1-\delta_a)^2+\frac{1}{g}(1-\delta_s)^2\right]-1}\Theta(\tilde \omega)$. Hence, 
even in the SILL there is a frequency range over which the Fermi edge behaves like that in a spinless LL.

Having completed the discussion for the Fermi edge singularity for a mobile hole at arbitrary magnetic field strength, we now turn to the case of an infinitely massive hole. It has been shown that for a spinless LL the infinitely massive core hole cuts the liquid in two and leads to a universal back scattering contribution to the exponent $\alpha$ [\onlinecite{Furusaki:prb97,Komnik:prb97,Prokofev:prb94,Gogolin:prl93}].  These boundary conditions kill the local density fluctuations and $\theta(\tau)=0$ at the impurity location.  Inspection of Eq.~\eqref{eq:C_psi_s} shows this boundary condition implies that only the $m=0$ term contributes, effectively eliminating all the spin incoherent factors $p^\sigma_m(-1)^m$.  As a result, {\em for all magnetic field values}, this problem maps onto the spinless LL case and $I(\omega)\propto \tilde \omega^{\frac{1}{g}\left(1-\delta_s\right)^2+1/8-1}\Theta(\tilde \omega)$, where the factor of $1/8$ is the universal backscattering contribution.  For the symmetry reasons mentioned in the introduction, there is a non-analytical behavior in the exponent when the infinite mass limit is taken for the hole\cite{Castella:prb96}. 

It should also be pointed out that the physical points just emphasized apply equally well to the single particle Greens function.  One must only set $\delta_a=\delta_s=\tilde E_h=0$. Thus, the tunneling density of states itself exhibits the crossover with respect to $\omega^*$.  Our numerics suggest that to clearly observe the spin incoherent effects, one must probe the lowest frequencies and to clearly observe the LL regime, one must probe frequencies approaching 1-10\% of $\omega_c$. We also note that in the SILL, there will be a peak in the tunneling density of states in finite field  at energies $\pm E_Z$ due to the absorption of energy between up and down states.  This is the SILL analog of the result in Ref.[\onlinecite{Shytov:prl05}].

In summary, we have studied the Fermi-edge singularity in a spin incoherent Luttinger liquid in a finite external magnetic field.  Two cases are distinguished--finite and infinite hole mass.  In the case of finite hole mass, the Fermi-edge singularity exhibits universal logarithmic corrections and the threshold exponent is independent of the hole mass, in contrast to a spin coherent Luttinger liquid.  The case of infinite hole mass can be directly mapped onto the equivalent spinless problem and no spin incoherent features appear.  Our predictions can be readily tested in quantum wires\cite{Calleja:prb95} at very low electron densities.

The author thanks L. Balents for a critical reading of the manuscript and M. Kindermann for pointing out an error in the estimate of $\omega^*$ in an earlier version. This work was supported by  NSF Grant numbers PHY99-07949, DMR04-57440, and the Packard Foundation.


\end{document}